\documentclass{article}
\usepackage{textgreek}
\usepackage{verbatim}
\usepackage{spconf,amsmath,graphicx}
\usepackage{multirow}
\usepackage{amsmath}
\usepackage{amssymb}
\usepackage{amsfonts}
\usepackage{color}
\usepackage{graphicx,marvosym}
\usepackage[dvipsnames]{xcolor}
\usepackage{tikz}
\usepackage{lipsum}

\title{A COMPARATIVE STUDY OF ESTIMATING ARTICULATORY MOVEMENTS FROM PHONEME SEQUENCES AND ACOUSTIC FEATURES}
\name{Abhayjeet Singh,  Aravind Illa, Prasanta Kumar Ghosh}
\address{Electrical Engineering, Indian Institute of Science (IISc), Bangalore-560012, India}

\begin{document}
 \ninept
\maketitle
\begin{abstract}
Unlike phoneme sequences, movements of speech articulators (lips, tongue, jaw, velum) and the resultant acoustic signal are known to encode not only the linguistic message but also carry para-linguistic information. While several works exist for estimating articulatory movement from acoustic signals, little is known to what extent articulatory movements can be predicted only from linguistic information, i.e., phoneme sequence. In this work, we estimate articulatory movements from three different input representations: R1) acoustic signal, R2) phoneme sequence, R3) phoneme sequence with timing information. While an attention network is used for estimating articulatory movement in the case of R2, BLSTM network is used for R1 and R3. Experiments with ten subjects' acoustic-articulatory data reveal that the estimation techniques achieve an average correlation coefficient of 0.85, 0.81, and 0.81 in the case of R1, R2, and R3 respectively. This indicates that attention network, although uses only phoneme sequence (R2) without any timing information, results in an estimation performance similar to that using rich acoustic signal (R1), suggesting that articulatory motion is primarily driven by the linguistic message. The correlation coefficient is further improved to 0.88 when R1 and R3 are used together for estimating articulatory movements.
\end{abstract}

\section{Introduction}
\label{sec:intro}
\begin{figure}[hbt]
   \centering
    \vspace{-.4cm}
   \centerline{\includegraphics[trim = 0mm 0mm 0mm 0mm, clip,width=5.15cm]{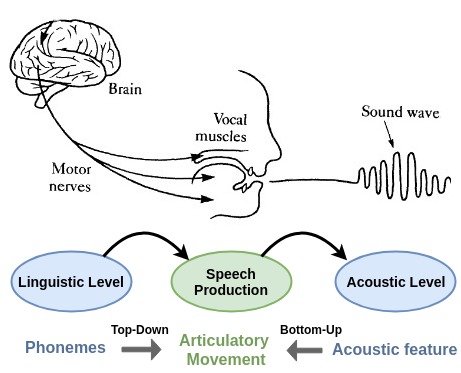}} 
    \vspace{-.4cm}
   	\caption{Top-down process involved in human speech production mechanism \cite{denes1993speech}}
   	\label{Speechchain}
 \vspace{-.3cm}
 \end{figure}

In speech production, articulatory movements provide an intermediate representation between neuro-motor planning (high level) and speech acoustics (low level) \cite{denes1993speech}. Fig. \ref{Speechchain} demonstrates the  top-down process involved in human speech production. Neuro-motor planning in the brain primarily aims to convey linguistic information (to express the thought) which are discrete abstract units. 
This information is passed through motor nerves to activate vocal muscles, which results in different temporally overlapping gestures of speech articulators (namely lips, tongue tip, tongue body, tongue dorsum, velum, and larynx), each of which regulates constriction in different parts of the vocal tract \cite{goldstein2003articulatory,neuro}. These articulatory gestures, in turn,  modulate the spectrum of acoustic signal, which results in the speech sound wave.
The acoustic features extracted from speech, embed para-linguistic information along with the linguistic content including speaker identity, age, gender, and emotion state. The para-linguistic information is often encoded in the  dynamics involved in the muscle coordination, articulatory timing, and vocal tract morphology of the speaker.
To estimate articulatory movements, in this work, we consider phonemes as a representative of high level information, and Mel-Frequency Cepstral Coefficeints (MFCC) as acoustic features. 
We also consider a representation by combining timing information with the phoneme sequence, which captures both linguistic and timing information but lacks the para-linguistic information and this could be treated as an intermediate representation between phonemes and MFCCs. From the acoustic features perspective, MFCC have been shown to carry maximal mutual information with articulatory features \cite{ghosh2010generalized,illa2019representation}.  Unlike estimating from MFCC (frame-to-frame estimation), estimating articulatory movements from discrete phonemes is a top-down approach and very challenging due to the absence of timing information. It is unclear to what extent we can estimate articulatory movement from the phonemes. 
In this work, we investigate on the accuracy with which the articulatory representations can be predicted from the phoneme sequence, and how it differs with respect to the acoustic features and phoneme sequence with timing information.

Knowledge about the articulatory position information along with the speech acoustics have shown to benefit applications like  language learning \cite{S2018,8462401}, automatic speech recognition \cite{ref8,ref9}, and speech synthesis \cite{Illa2019,ling2009integrating} tasks. A rich literature exists in estimating articulatory movements from acoustic features of speech which is typically known as acoustic-to-articulatory inversion (AAI). Various approaches were proposed including 
Gaussian Mixture Model (GMM) \cite{ref12}, Hidden Markov Model (HMM) \cite{ref14}, and neural network \cite{ref13, wu2015acoustic}. The state-of-art performance is achieved by bidirectional long short-term memory networks (BLSTM)\cite{illa2018low,BLSTM}.

On the other hand, there have been few attempts for estimating articulatory movements from text or phonological units. These attempts typically deployed the techniques from speech synthesis paradigm, using HMM \cite{HMMP2A} and BLSTM \cite{BLSTMP2A}. In fact, these works compared their respective performance with that from AAI. Works in \cite{HMMP2A,BLSTMP2A} reported a significant drop in phonological feature based performance compared to that using acoustic features (AAI model). The main reason for the drop in the performance from acoustic features to phonemes could be due to the limitations of the duration model with HMM \cite{HMMP2A,BLSTMP2A}.

With advancements in the neutral network based modelling approaches, it has been shown that attention networks are able to capture the duration model more accurately, and achieve the state-of-the-art performance in speech synthesis applications \cite{taco2}. In this work, we deploy the tacotron \cite{taco2} based speech synthesis approach for phonemes to articulatory movement prediction. 
Since, the tacotron based model needs large amount of data to learn the alignments and the phonemes-to-articulatory movements mapping, we deploy generalized model training using ten subjects' training data and fine-tuning approach for each subject as done for AAI in \cite{illa2018low}. 
A systematic comparison of articulatory movement prediction using different features reveals that a correlation coefficient of 0.81 between predicted and original articulatory movements is obtained when phoneme sequence is used for prediction, without any timing information. On the other hand, the acoustic features based prediction achieves a correlation coefficient of 0.85 indicating that articulatory movements are primarily driven by linguistic information.


\vspace{-0.3cm}
\section{Data set}
\begin{figure}[htb]
   \centering
    \vspace{-.4cm}
   \centerline{\includegraphics[trim = 7mm 214mm 35mm 7mm, clip, width=5cm]{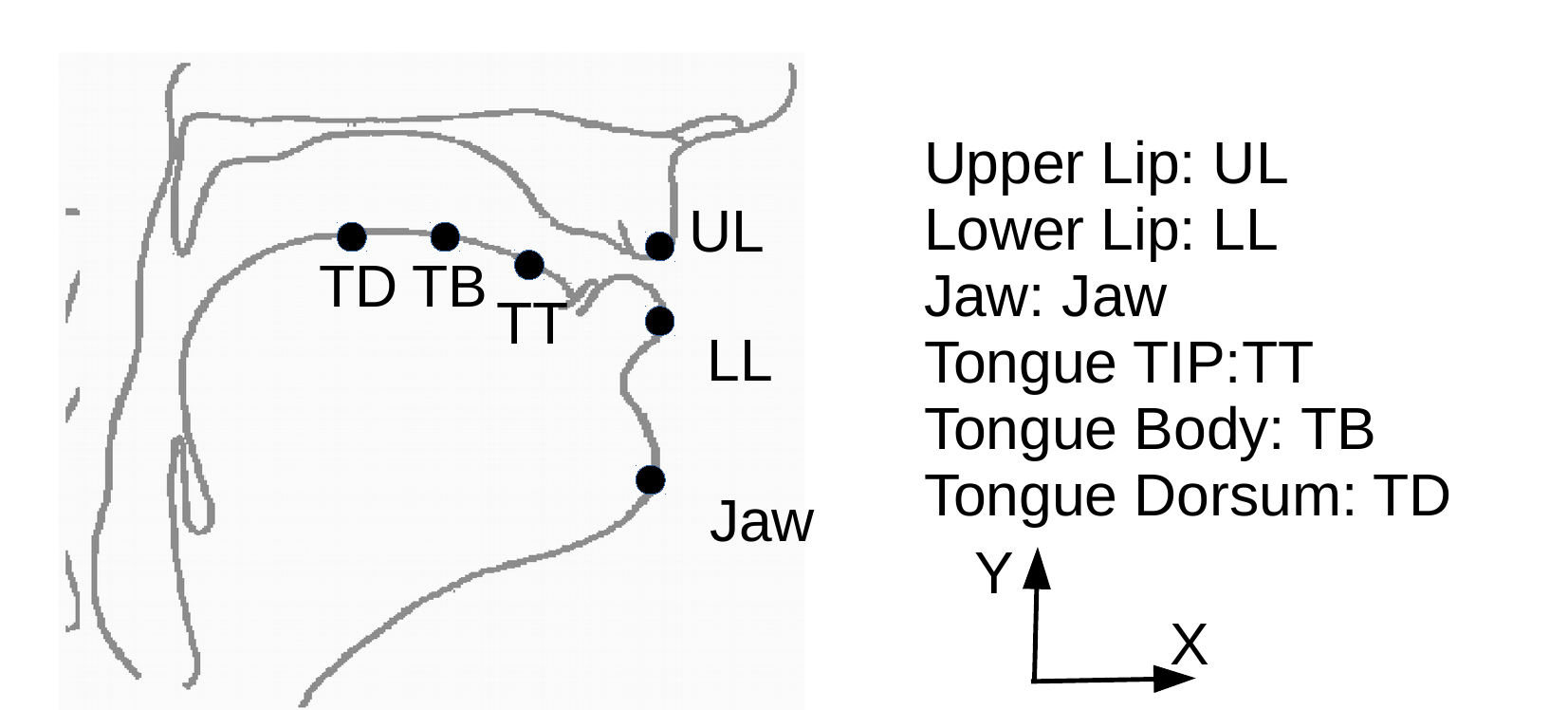}} 
    \vspace{-.4cm}
   	\caption{Schematic diagram indicating the placement of EMA sensors \cite{illa2018low}.}
   	\label{EmaSetUp}
 \vspace{-.3cm}
 \end{figure}

In this work, we consider a set of 460 phonetically balanced English sentences from MOCHA-TIMIT corpus as the stimuli for data collection from a group of 10 subjects comprising 6 males (M1, M2, M3, M4, M5, M6) and 4 females (F1, F2, F3, F4) in the age group of 20-28 years. All the subjects are native Indians with proficiency in English and reported to have no speech disorders in the past. All subjects were familiarized with the 460 sentences to obviate any elocution errors during recording. For each sentence, we simultaneously recorded audio using a microphone \cite{EM9600} and articulatory movement data using Electromagnetic articulograph (EMA) AG501 \cite{AG501}. EMA AG501 has 24 channels  to measure the horizontal, vertical, and lateral displacements and angular orientations of a maximum of 24 sensors. The articulatory movement was collected with a sampling rate of 250Hz. The sensors were placed following the guidelines provided in \cite{optimal}. A schematic diagram of the sensor placement is shown in Fig. \ref{EmaSetUp}.

As indicated in Fig. \ref{EmaSetUp}, we used 6 sensors which were glued on different articulators, viz. Upper Lip (UL), Lower Lip (LP), Jaw, Tongue Tip (TT), Tongue Body (TB), and Tongue Dorsum (TD). 
Each of these six sensors captured the movements of articulators in 3D space. 
Additionally, we also glued two sensors behind the ears for head movement correction.
In this study we consider the movements only in the midsagittal plane, indicated by X and Y directions in Fig. \ref{EmaSetUp}. Thus, we have twelve articulatory trajectories denoted by $UL_x$, $UL_y$, $LL_x$, $LL_y$, $Jaw_x$, $Jaw_y$, $TT_x$, $TT_y$, $TB_x$, $TB_y$, $TD_x$, $TD_y$. Before placing the sensors, the subjects were made to sit comfortably in the EMA recording setup. The subjects were given sufficient time to get used to speaking naturally with the sensors attached to different articulators. Manual annotations were performed to remove silence at the start and end of each sentence.


\vspace{-0.3cm}
\section{Proposed Approach}
In this section, we first present the proposed approach for phoneme-to-articulatory prediction, and then describe a BLSTM based AAI approach, followed by a training scheme to overcome the limitation of scarcity of acoustic-articulatory data.

\begin{figure}[htb]
   \centering
   \vspace{-.28cm}
   \centerline{\includegraphics[trim = 25mm 16mm 10mm 20mm, clip,width=6.5cm]{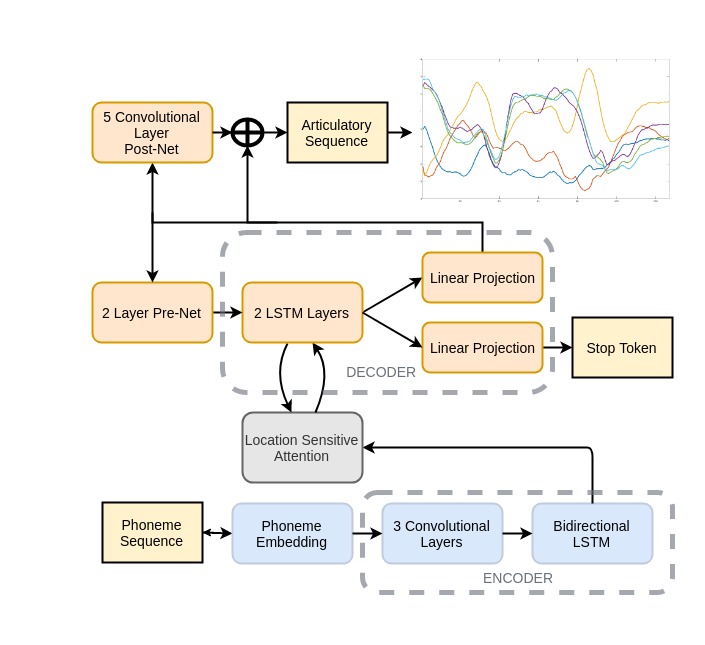}}
    \vspace{-.35cm}
    \caption{Block diagram for phoneme sequence to articulatory movement prediction approach \cite{taco2}.}
    \label{blk_dig}
    \vspace{-.6cm}
\end{figure}

\textit{Articulatory movement estimation using attention network:}
In this work, we deploy the state-of-the-art speech synthesis model, tacotron architecture \cite{taco2} to model duration information for articulatory movement estimation from phonemes.
There are three major components in tacotron model, encoder, attention and decoder as shown in Fig. \ref{blk_dig}.
Encoder takes discrete sequences of phonemes 
$P=\{p_1,p_2,..p{_N}\}$ 
as an input and maps it to the hidden states $E=\{e_1,e_2,..e_N\}$, which acts as an input to the attention.
The attention network models the time alignment between the encoder and decoder hidden states. The decoder hidden states, $D=\{d_1,d_2,..d_T\}$, are utilized to generate the articulatory movement trajectories over time.

Each phoneme is represented by a 40-dimensional one-hot encoded representation, which are fed to a 512-dimensional embedding layer, followed by a stack of 3 convolution layers and a BLSTM layer with 512 hidden units, which make up the encoder block of the model as depicted in Fig. \ref{blk_dig}. Attention network is a location-sensitive attention network which iterates over previous decoder output ($d_{t-1}$) and attention weights ($\alpha_{n,t-1}$) and
all the encoder hidden states ($E$).
The attention mechanism is governed by the equations below \cite{ASRattention}:
\begin{equation}
 \alpha_{n,t}=\sigma (\text{score}(d_{t-1},\alpha_{n,t-1},E))
\end{equation}
 \begin{equation}
 \text{score:}\ s_{n,t}=w^T \text{tanh}(Wd_{t-1}+Ve_n+Uf_{j,t}+b)
\end{equation} 
\begin{equation}
    \text{Context vector: } g_t=\sum_{j=1}^N\alpha_{j,t}e_{j}
\end{equation}
where $\alpha_{n,t}$ are \textit{attention weights} and $n \epsilon \{1,2..N\}$ and parameters of attention network are denoted by
$W$, $V$, $U$ weight matrices and $w$ and $b$ denotes the projection and bias vectors, respectively. In the Eq.(2), $f_{j,t}$ is computed by $f_t=F*\alpha_{t-1}$, which incorporates location  awareness to the attention network \cite{ASRattention}, where $F$ is a weight matrix and $\alpha_{t-1}$ is the set of previous time-step alignments.
In Eq.(2), attention scores are computed as a function of encoder outputs ($e_n$) and previous attention weights ($\alpha_{n,t-1}$) and decoder output ($d_{t-1}$), which are further normalized using \textit{softmax} 
to obtain attention weights.
These obtained attention weights are utilized to compute fixed context vector as described by Eq.(3).

 The decoder consists of two uni-directional LSTM layers with 1024 units, followed by a linear projection layer to predict the articulatory movements as shown in Fig. \ref{blk_dig}. 
The decoder computes the final output ( $d_t \sim Decoder(d_{t-1},g_{t})$) 
from the previous state output ($d_{t-1}$) and attention context vector ($g_t$). 
To compute $d_{t-1}$, decoder's previous output is transformed by a two layered fully-connected network with 256 units (Pre-Net).
The decoder hidden state outputs are further projected using two linear layers, one for articulatory sequence prediction and other to predict end of the sequence. The predicted articulatory trajectories are passed through a 5-layer convolutional Post-Net which predicts a residual to add to the prediction to improve the overall reconstruction. Each layer in the Post-Net consists of 512 filters with a dimension of $5\times1$ followed by a batch normalization layer. Tanh activation is used at the end of each layer except for the final layer.
For end sequence prediction, the decoder LSTM output and attention context are concatenated and projected down to a scalar and then passed through sigmoid activation to predict the probability that the output sequence has completed. This ``Stop Token" prediction is used during inference to allow the model to dynamically determine when to terminate generation instead of always generating for a fixed duration.
 
\textit{Articulatory movement estimation using BLSTM:}
Acoustic features and phoneme features with timing information have one-to-one correspondence between the input and output (articulatory position information) features, thereby implicitly encode timing information.
Therefore, we do not need to model timing information explicitly for articulatory movement prediction in such cases.
Deep recurrent neural networks architecture, namely BLSTM network has been shown to perform well in capturing context and smoothing characteristics  and  achieves  the  state-of-art  AAI  performance \cite{illa2018low,BLSTM}.
So, in this work, we deploy a BLSTM neural network based method to estimate articulatory movements from acoustic and phoneme features with timing information.
These input features are fed to three consecutive BLSTM layers with 256 units and at the output, we use a time distributed dense layer as a linear regression layer.

\textit{Generalized model and fine-tuning:}
Typically neural network models demand large amount of training data to achieve better performance.
To overcome the scarcity of articulatory data to train a subject specific model, we deploy the training approach proposed in \cite{illa2018low}.
At the first step, we pool the training data from all the subjects and train a generic model to learn the mapping from input features to the target articulatory trajectories. In the second stage, we fine tune the generic model weights with respect to the target speaker, to learn speaker specific models.

\vspace{-0.3cm}
\section{Experimental Setup}
\textit{Data pre-processing and feature computation}:
Articulatory data is first low-pass filtered to remove high frequency noise incurred during measurement process. The cutoff frequency of low-pass filter is 25Hz, which preserves the smoothly varying nature of the articulatory movement \cite{ghosh2010generalized}.
Further, we perform sentence wise mean removal and variance normalization along each articulatory dimension.
The recorded speech is down-sampled from 48kHz to 16kHz, following which a forced  alignment is performed to obtain phonetic transcription using Kaldi \cite{povey2011kaldi}. The resultant phonetic transcription of the dataset consists of 39 ARPABET symbols.
For experiment with phonemes,
we use 39 discrete phonemes plus a start token, which are encoded as a 40-dim one-hot encoded vector, denoted as ``PHN". We also incorporate the timing information to phoneme sequence using phonetic boundaries obtained from the forced alignment. In this feature set, we replicate the one-hot vector for every 10 msec in the entire duration of the corresponding phonemes, which we denote  by ``TPHN". For acoustic features, we compute a 13-dim Mel-frequency cepstrum coefficients (MFCC), which are shown to be optimal for the AAI task \cite{ghosh2010generalized,illa2019representation}.
This TPHN feature set carries both linguistic information and timing information, and, hence, it works as an intermediate representation carrying information between MFCC and PHN. We also experimented with the concatenation of MFCC and TPHN to observe if there is any benefit of providing linguistic information explicitly along with MFCC.
A summary of input features with respect to the hypothesized possible information they encode and the models used to estimate articulatory movements are reported in Table \ref{methods}.


\begin{table}
\centering
\caption{Summary of input features, corresponding encoded information and models used for articulatory movement estimation.}
\vspace{.1cm}
\label{methods}
\resizebox{0.52\textwidth}{!}{\begin{tabular}{|c|c|c|cc}
\cline{1-3}
\textbf{Input Features}                                                & \textbf{Encoded Information}                                                 & \textbf{Model} &  &  \\ \cline{1-3}
\begin{tabular}[c]{@{}c@{}}Phonemes Sequence (PHN)\end{tabular}      & linguistic                                                                   & Attention     &  &  \\ \cline{1-3}
\begin{tabular}[c]{@{}c@{}}Time Aligned Phonemes (TPHN)\end{tabular} & linguistic+timing                                                            & BLSTM          &  &  \\ \cline{1-3}
MFCC                                                                   & \begin{tabular}[c]{@{}c@{}}linguistic+para-linguistic\\ +timing\end{tabular} & BLSTM          &  &  \\ \cline{1-3}
MFCC+TPHN                                                              & \begin{tabular}[c]{@{}c@{}}linguistic+para-linguistic\\ +timing\end{tabular} & BLSTM          &  &  \\ \cline{1-3}
\end{tabular}}
\vspace{-.5cm}
\end{table}

\textit{Model training and evaluation procedure:}
The recorded acoustic-articulatory data from 460 sentences are divided into train ($80\%$), validation ($10\%$), and test ($10\%$) sets, for each subject. 
For BLSTM network training, we perform zero padding for all the sentences in train and validation set to obtain a fixed length sequence of 400 frames (4sec) with a batch size of 25. 
We choose root mean squared error (RMSE) and correlation coefficient (CC) as evaluation metrics to assess the performance the articulatory movement estimation technique.
The RMSE and CC between the original and estimated trajectories are computed for each articulator separately \cite{ghosh2010generalized,illa2018low}. 
Note that the RMSE is computed on articulatory trajectories which are mean and variance normalized. So RMSE does not have any units.


For experiments with PHN, we perform teacher forcing approach \cite{taco2} while training attention network, by iterating over the decoder hidden states output $d_{t-1}$. We pass previous ground-truth articulatory position information as an input to attention and decoder instead of $d_{t-1}$. 
During testing, we perform DTW alignment using Euclidean distance metric between the predicted and estimated articulatory trajectories, and then compute the RMSE and CC to assess the model performance. We use RMSE as an objective measure for learning weights for all the models.

Three types of training were done for all features: Subject-Dependent, Generic, and Fine-Tuned. In the first case, the models were trained for each subject separately, giving 10 models for 10 subjects, whereas in the second case, a single generic model is trained on all 10 subjects pooled together. 
And in the third case we trained the models for each subject separately, by fine-tuning the generic model with subject specific data. All the experiments for BLSTM were performed using Keras \cite{Keras} and Tensorflow backend \cite{TF}. Experiments with attention network were performed using NVIDIA open source code repository \cite{NVIDIAOpenSourceCode}.
\vspace{-0.3cm}
\section{Results and Discussion}
In this section, we first present the results of the generalized model and fine-tuning approach to estimate articulatory movements. Then we compare the performance with different features (PHN, TPHN, MFCC) in a subject specific 
manner.

\vspace{-0.2cm}
\subsection{Generalized model and fine-tuning}
Table \ref{PrefComp}, reports the performance with different training approaches across features. Interestingly, we observe that in all the cases fine-tuning of a model performs better than the subject-dependent model. This implies that pooling the data from all subjects helps in learning generic mapping across multiple subjects, and fine-tuning with speaker specific training data improves the speaker specific mapping. On the other-hand, while comparing the performance of the generic model with the subject dependent model, we observe that TPHN performance drops in generic model. This could be due to the lack of speaker specific para-linguistic information in the input features, unlike MFCC which encodes speaker information and enables BLSTM to learn mappings for multiple subjects by a single model \cite{illa2018low}.
Although speaker information lacks when PHN is used as the input feature, the performance using subject-dependent model is worse than that using generic model. This is primarily due to scarcity of the training data for the complex attention network.
The relative improvements from generic to fine-tuned model across the PHN, TPHN, MFCC, and MFCC+TPHN are 18.53, 9.3, 0.5, and 0.8$\%$, respectively. Unlike MFCC, the greater improvements in PHN could be due to lack of para-linguistic information conveying speaker information for generic model with pooled data, which could lead to the coarse duration model across multiple subjects which, when fine-tuned with individual subject's data, becomes more subject specific. We also perform experiments with MNGU corpus to compare tacotron based PHN model with HMMs \cite{HMMP2A}. We obtain 0.78 CC with tacotron model fine-tuned with MNGU corpus, while \cite{HMMP2A} reports 0.60 CC with HMM based approach.

\begin{table}
\centering
 	\renewcommand\tabcolsep{1pt}
 	\vspace{-.5cm}
 	 \caption{Performance comparison across different features. Numbers in brackets denote the standard deviation across all test sentences.}
 	 \label{PrefComp}
	 \def\arraystretch{1}
\resizebox{0.48\textwidth}{!}{\begin{tabular}{|p{4.3em}|c|c|c|c|c|c|c|c|}
\hline
\multirow{2}{*}{Training} & \multicolumn{2}{c|}{PHN}                                   & \multicolumn{2}{c|}{TPHN}                                   & \multicolumn{2}{c|}{MFCC}                                   & \multicolumn{2}{c|}{MFCC+TPHN}
\\ \cline{2-9} 
                          & RMSE & CC                                                  & RMSE & CC                                                  & RMSE & CC                                                  & RMSE & CC                                                   \\ \hline
Subject-Dependent         & \begin{tabular}[c]{@{}c@{}}2.04\\ (0.109)\end{tabular}     & \begin{tabular}[c]{@{}c@{}}0.33\\ (0.031)\end{tabular} & \begin{tabular}[c]{@{}c@{}}1.243\\ (0.087)\end{tabular}      & \begin{tabular}[c]{@{}c@{}}0.808\\ (0.033)\end{tabular} & \begin{tabular}[c]{@{}c@{}}1.116\\ (0.095)\end{tabular}     & \begin{tabular}[c]{@{}c@{}}0.844\\ (0.025)\end{tabular}             & \begin{tabular}[c]{@{}c@{}}1.05\\ (0.086)\end{tabular}     & \begin{tabular}[c]{@{}c@{}}0.87\\ (0.024)\end{tabular} \\ \hline
Generic                   & \begin{tabular}[c]{@{}c@{}}1.48\\ (0.098)\end{tabular}     & \begin{tabular}[c]{@{}c@{}}0.68\\ (0.043)\end{tabular} & \begin{tabular}[c]{@{}c@{}}1.44\\ (0.108)\end{tabular}     & \begin{tabular}[c]{@{}c@{}}0.74\\ (0.046)\end{tabular} & \begin{tabular}[c]{@{}c@{}}1.107\\ (0.091)\end{tabular}     & \begin{tabular}[c]{@{}c@{}}0.849\\ (0.023)\end{tabular}             & \begin{tabular}[c]{@{}c@{}}1.01\\ (0.083)\end{tabular}     & \begin{tabular}[c]{@{}c@{}}0.877\\ (0.022)\end{tabular}\\ \hline
Fine-Tuned               &  \begin{tabular}[c]{@{}c@{}}1.18\\ (0.113)\end{tabular}    & \begin{tabular}[c]{@{}c@{}}0.806\\ (0.039)\end{tabular} & \begin{tabular}[c]{@{}c@{}}1.239\\ (0.084)\end{tabular}     & \begin{tabular}[c]{@{}c@{}}0.815\\ (0.033)\end{tabular} & \begin{tabular}[c]{@{}c@{}}1.090\\ (0.088)\end{tabular}     & \begin{tabular}[c]{@{}c@{}}0.854\\ (0.024)\end{tabular}             & \begin{tabular}[c]{@{}c@{}}0.99\\ (0.085)\end{tabular}     & \begin{tabular}[c]{@{}c@{}}0.884\\ (0.021)\end{tabular}\\ \hline
\end{tabular}}
\vspace{-.4cm}
\end{table}

 \begin{figure}[htb]
   \centering
   \vspace{-.3cm}
   \centerline{\includegraphics[trim = 34mm 200mm 40mm 1mm, clip,height=3cm,width=9cm]{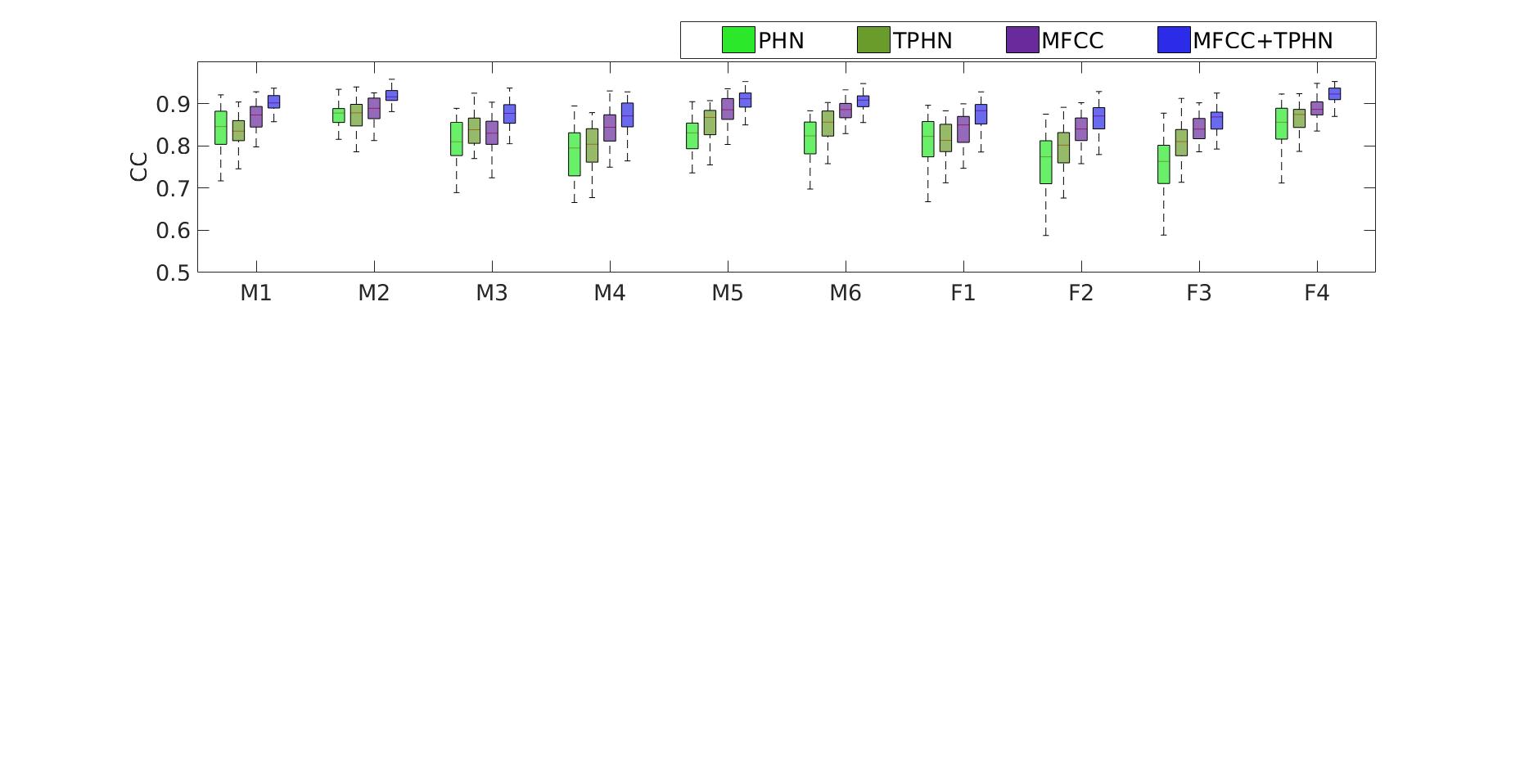}}
    \vspace{-.3cm}
    \caption{CC across all articulators for each speaker}
    \label{subs}
    \vspace{-.6cm}
 \end{figure}
 
\vspace{-0.2cm}
\subsection{Comparsion of performance across features}
We compare the performance of models with respect to input features, namely PHN, TPHN, MFCC, and TPHN+MFCC (concatenation of TPHN and MFCC). Fig. \ref{subs}, illustrates the box plot of CC using different input features, where x-axis represents different subjects and y-axis indicates the CC across all the articulators. While comparing the performance of PHN with TPHN, we observe that there is no significant difference ($p < 0.01$) in performance for six subjects, namely M1, M2, M4, M6, F1, F2 and F4; which indicates that the timing information could be recovered from phoneme sequence and articulatory trajectories using attention network. This helps it to perform similar to the BLSTM model with TPHN features, which explicitly provides timing information. We observe that MFCC outperforms both PHN and TPHN features, this could be due to the fact that articulatory information is maximally preserved when speech acoustic signal is processed by auditory filters such as mel-scale \cite{ghoshAudprocessing,illa2019representation}. Experiments reveal that fusing the TPHN with MFCC features  indicated by MFCC+TPHN, results in the best performance among all the features.

 
  \begin{figure}[htb]
   \centering
   \vspace{-.2cm}
   \centerline{\includegraphics[trim = 0mm 38mm 0mm 33mm,height=8.3cm,width=9cm]{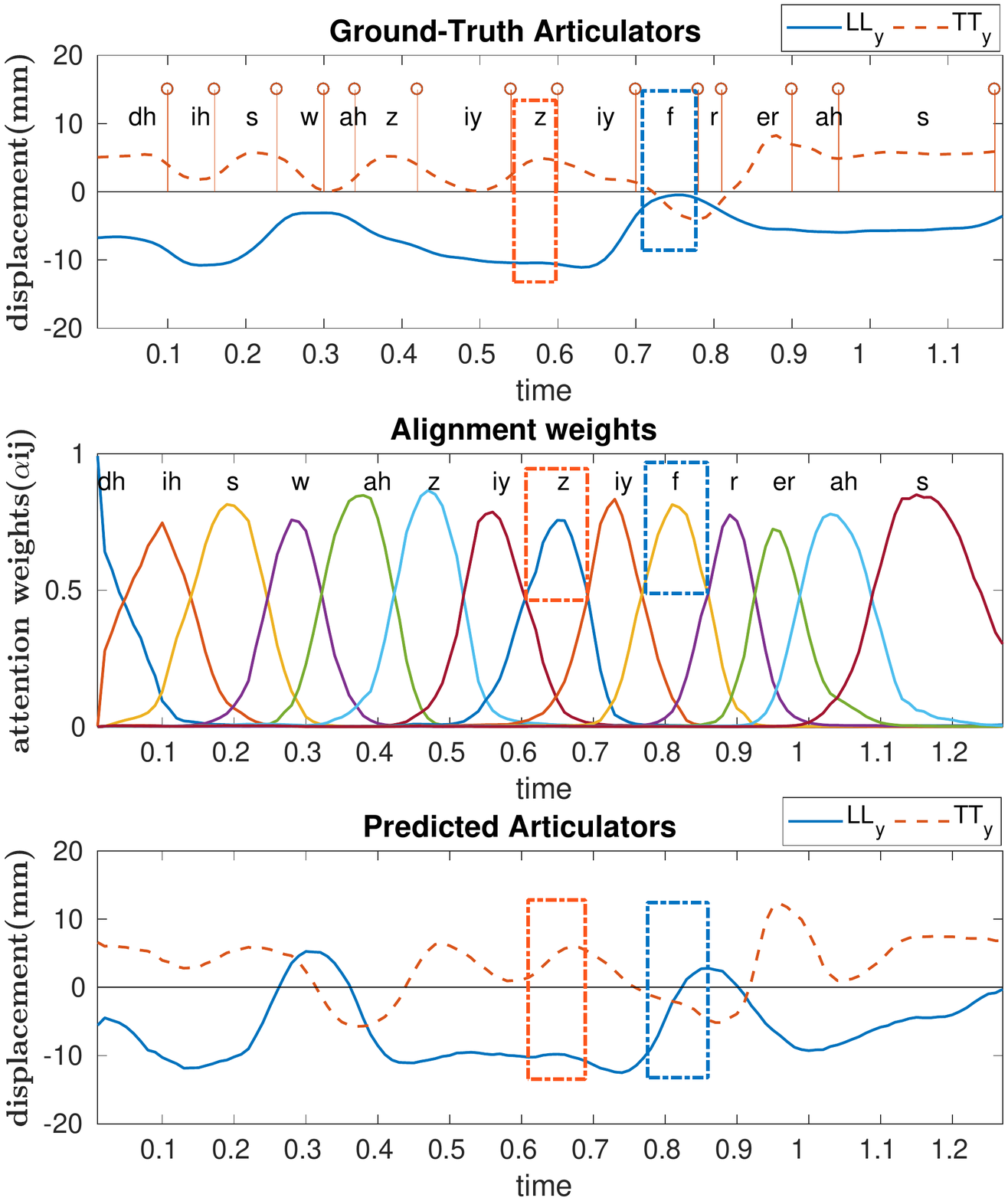}}
    \vspace{-.4cm}
    \caption{Relating attention weights to transitions in predicted articulatory trajectory in comparison with the corresponding ground truth for each phoneme label}
    \label{time_bound}
    \vspace{-.6cm}
 \end{figure}

\subsection{Illustration of attention weights ($\alpha$)}
To illustrate the attention weights learned for PHN to articulatory movement mapping in the fine-tuned model, we consider an example utterance ``This was easy for us".
 In Fig. \ref{time_bound}, in the top subplot, we plot the ground-truth articulatory trajectories for lower lip ($LL_y$) and tongue tip ($TT_y$) and the corresponding phonetic boundaries are indicated by vertical lines.
 The attention weights are illustrated in the second subplot, where their corresponding phoneme labels are indicated on top of each weight profile. The estimated articulatory trajectories are plotted in the last subplot.
Let us consider the attention weight corresponding to phoneme `z'  in ``easy". When the attention network provides a greater weight to the hidden state corresponding to phoneme `z' (indicated by rectangular box), there is a tongue tip constriction in the predicted trajectories.
Similar trend is observed with attention weight for the phoneme ``f" in the word ``for'', where we observe lip closures as a peak in lower lip vertical ($LL_y$) movement. We observe that the trend in the original and predicted articulatory trajectories are similar. 

\vspace{-0.3cm}
\section{Conclusion}

In this work, we proposed phoneme-to-articulatory movement estimation using attention networks. We experimentally showed that, with phoneme sequence without any timing information, we achieve an estimation performance which is identical to that using timing information. This implies that attention networks are able to learn the timing information to estimate articulatory movements. Experiments performed with different features, reveal that MFCC concatenated with TPHN features achieve the best performance in estimating articulatory movements. In future, we plan to utilize the estimated articulatory movements in speech synthesis task and in developing audio-visual speech synthesis systems.\\


\vspace{-0.3cm}
\noindent\rule{4cm}{0.4pt}\\
{\footnotesize Authors thank all the subjects for their participation in the data collection. We also thank the Pratiksha Trust and the Department of Science and Technology, Govt. of India for their support in this work.}



\bibliographystyle{IEEEbib}
\bibliography{refs}

\end{document}